\documentclass[preprint,aps]{revtex4}
\usepackage{graphicx}
\begin{document}
\title{P-wave Radial distributions of a Heavy-light meson on a lattice}

\author{UKQCD Collaboration, A.M. Green
\thanks{email: anthony.green@helsinki.fi}}
\affiliation{Helsinki Institute 
of Physics, P.O. Box 64, FIN--00014 University of Helsinki, Finland}
\author{J. Koponen}
\affiliation{Department of Physical Sciences and Helsinki Institute of
Physics\\
P.O. Box 64, FIN--00014 University of Helsinki, Finland}
\author{C. Michael}
\affiliation{Department of Mathematical Sciences, University of Liverpool,
 L69 3BX, UK}

\date{\today}

\begin{abstract}
This is a follow-up to our earlier work  for the charge (vector)
and matter (scalar) distributions for S-wave states in a heavy-light
meson,
where the heavy quark is static and the light quark has a mass about
that
of the strange quark. The calculation is again carried out with
dynamical
fermions on a $16^3\times 24$ lattice with a lattice spacing of about 0.14 fm.
It is shown that several features of the S- and P-wave distributions are
in
qualitative agreement with what one expects from a simple one-body Dirac
equation interpretation.
\end{abstract}

\maketitle

\section{Introduction}
Experimentalists are often able to tell us several properties of a given  meson,
such as its energy, width and angular momentum. However, usually they can not
tell us the structure of the meson. For example, with $B_s$ states ---
the topic of interest here --- they can
not say whether these states are $b\bar{s}$, $b\bar{s}u\bar{u}$ or
$B K$. Unfortunately, when theoreticians try to describe $B_s$
states, they have to decide beforehand the state structure to be used in  some
model, which often has sufficient freedom to fit the data with any of
the possible structures. In an attempt to clarify the Experiment
$\longleftrightarrow$ Theory comparison, we suggest the use of lattice QCD. In
principle, lattice QCD should give us all we need to know about $B_s$
states. However, in practice, the results need to be corrected for the
lattice spacing ($a$), finite lattice size ($L$)  and quark mass
($m_q$) effects --- but these are usually under control and with the
advent of more computer resources they will decrease in importance.

The strategy followed here is to concentrate on the simplest of quark
states, namely, the $Q\bar{q}$ system, where $Q$ is an infinitely heavy quark
({\it i.e.} static) and $q$ is a quark with about the strange quark mass.
This system is sufficiently simple to enable state-of-the-art lattice
 calculations to generate much more ``data'' than can be achieved by
direct experiment. This data consists of the ground state energies of
S-, P- and D-wave states and also the spin average F-wave energy \cite{PRD69}.
Not only are the ground
state energies extracted, but also those of the corresponding excited states
containing at least one radial node. In addition, the vector (charge) and
scalar (matter) radial distributions of these states can be measured. This is
an abundance of data, far beyond what has been done experimentally in
the $B_s$ meson. Given all this data the challenge is now for theorists to
make models  to explain it. In this quest there are two simplifications.
Firstly, since $Q$ is static, the system is essentially reduced to a
one-body problem involving only the light quark $q_s$. Secondly,
in the lattice calculations, since the energies and radial distributions
are extracted from $Q\bar{q}$ correlations propagating in Euclidean
time, it is expected that the resultant states are indeed $Q\bar{q}$
states with little contamination from other possible multiquark
components. Support for this expectation (hope) is seen in Fig. 2 of
Ref.~\cite{PRD63}. There a $Q\bar{Q}$ correlation generates a linearly
rising potential for interquark distances far larger than expected 
{\it i.e.}
way beyond  where $(Q\bar{q})(\bar{Q}q)$ configurations should appear through
string breaking. To see this effect the explicit introduction
of $(Q\bar{q})(\bar{Q}q)$ correlations is needed \cite{CMPP}.

The ``data base'' for  properties of $Q\bar{q}$ states measured on lattices
is so far incomplete. In Ref.~\cite{EPJ28} we concentrated on the
S-wave energies and radial distributions. The latter showed two distinct
features. Firstly, the excited state distributions exhibited nodes
as expected.
Secondly, the charge distribution ($x_C(R)$) was of longer range
than the matter distribution ($x_M(R)$) but with $x_C(0)\approx x_M(0)$.
This is readily explained by the one-body Dirac equation, since there
$x_C(R)=G(R)^2+F(R)^2$ and $x_M(R)=G(R)^2-F(R)^2$, where for S-waves
$G(0)\gg F(0)$ with $G(R)$ and $F(R)$ becoming comparable for large $R$.
In Ref.~\cite{PRD69} the data base was extended to include the energies
of the excited states $P_{+/-}$ and $ D_{+/-}$ and also the spin averaged
combinations $D_{+-}$ and $F_{+-}$. Here $P_-$ is the $P$-wave state
with $j_q=1/2$, since the spin of the $Q$ does not play a dynamical
role.
In this note we return to measuring the charge and matter distributions
as in Ref.~\cite{EPJ28}, but concentrate on the $P_-$-state and its 
excitations. This state is of particular interest, since --- as shown in
Fig.~4 of Ref.~\cite{PRD69} --- the indications are that the predicted 
$B_s(0^+)$ is below the $BK$-threshold and so should be very narrow as was found
for the $c\bar{s}$ counterpart.
The outcome is seen in Figs.~1 for the ground
state distributions $x^{11}$ and the off-diagonal distributions
between the ground state and the first excited state $x^{12}$. 

The
latter show single nodes as expected from a first excited state.
However, $x^{11}$ shows two features which at first sight seem 
surprising: 

1) The  distributions are {\it finite} at $R=0$, even though they 
are $P$-waves.

2) The matter distribution has a {\it node}, even though it involves only
the ground state.
 
\noindent But again this is precisely what one expects from solutions of the Dirac
equation, where for the $P_-$-state both $G$ and $F$ are {\it non-zero} at
$R=0$ and, furthermore, the ``small'' component $F$ can be {\it larger}
than $G$ at small $R$.
In Figs.~2 the above data, now with a factor of $R^2$ included, are
compared with the Dirac distributions using a quark mass of 100 MeV and
an interquark potential $V=-a/R+bR$, where $a=0.6$ and $b$=1.3 GeV/fm.
These three parameters are very sensible and were not tuned to get an
optimal fit. 

 \begin{figure}
 \begin{center}
\includegraphics*[width=0.98\textwidth]{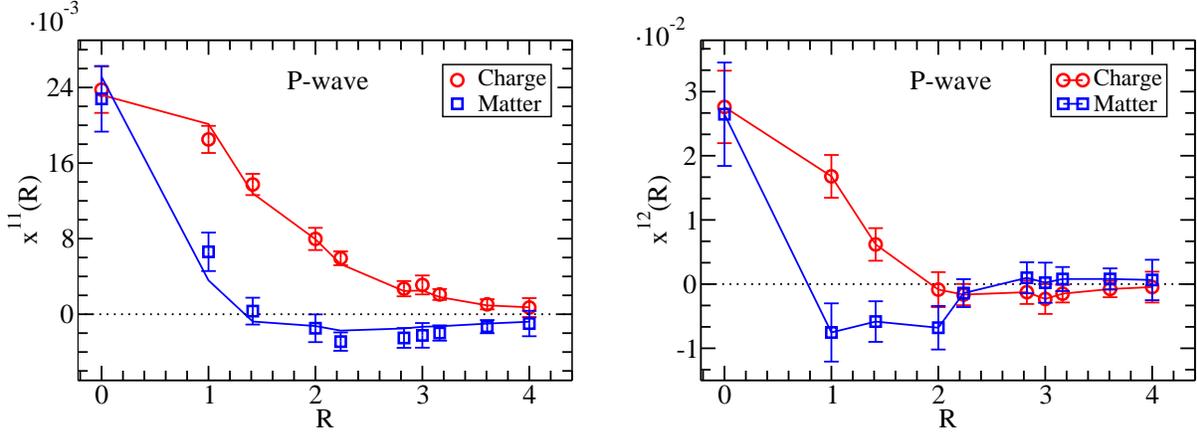}
\end{center}
  \caption{The $P_-$ charge and matter distributions for
$x^{11}(R)$ and $x^{12}(R)$.} 
\end{figure}

\begin{figure}
 \begin{center}
\includegraphics*[width=0.95\textwidth]{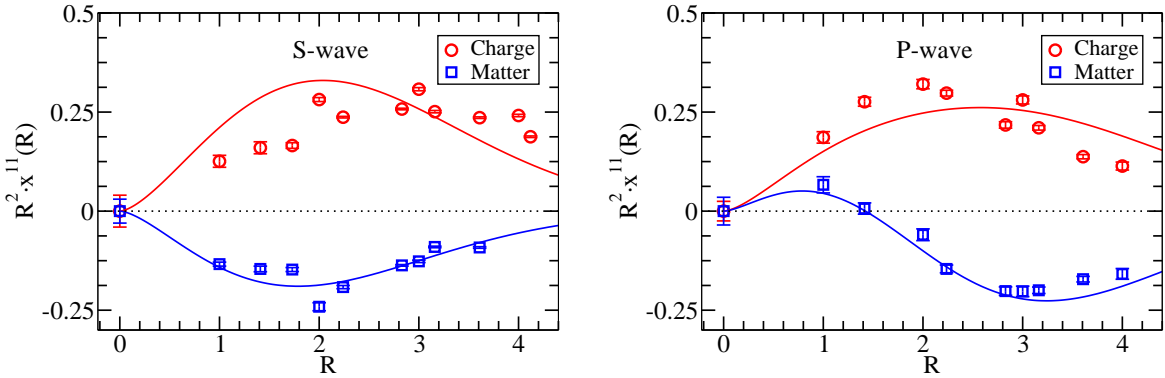}
\end{center}
 \caption{The $S$ and $P_-$ radial distributions from Fig.~1
compared with Dirac equation fits.}         
\end{figure}

\section{Conclusions}
The S- and P-wave charge and matter distributions in Figs.~2 suggest that they
can be understood qualitatively in terms of the one-body Dirac equation.
The challenge is now to see to what extent there is an analogous 
{\it quantitative}  description of the energies and distributions of all the
S-, P-, D- and F-wave states --- both ground and excited, when they
become available. It is possible
that the  strategy used in studying the $NN$-potential is appropriate,
namely, to first concentrate on the higher partial waves and so avoid or
reduce complications, such as the effect of form factors needed to regulate the 
one-gluon-exchange potential and also instanton-induced interactions, 
that enter at small values of $R$.

The authors wish to thank the Center for Scientific Computing in Espoo,
Finland and the ULGrid project of the University of Liverpool for making
available ample computer resources. Two of us, AMG and JK, acknowledge
support by the Academy of Finland contract 54038 and
the EU grant HPRN-CT-2002-00311 Euridice.
JK thanks the Finnish Cultural Foundation and the Magnus Ehrnrooth
Foundation for financial support.


\end{document}